\documentclass[11pt]{article}
\usepackage{graphicx}
\usepackage[margin=1.25in]{geometry}
\usepackage[usenames,dvipsnames]{color}
\usepackage{url}
\usepackage[colorlinks = true,
            linkcolor = blue,
            urlcolor  = blue,
            citecolor = blue,
            anchorcolor = blue]{hyperref}

\usepackage[export]{adjustbox}
\usepackage{wrapfig}
\usepackage{cite}
\usepackage{float}

\def\Title#1{\begin{center} {\LARGE #1 } \end{center}}
\def\Author#1{\begin{center}{ \sc #1} \end{center}}

\newenvironment{Abstract}{\begin{quotation} \begin{center}
                       ABSTRACT
     \end{center}\bigskip  }{\end{quotation}}

\def\beq{\begin{equation}}
\def\eeq#1{\label{#1}\end{equation}}
\def\eeqn{\end{equation}}

\newenvironment{Eqnarray}%
   {\arraycolsep 0.14em\begin{eqnarray}}{\end{eqnarray}}
\def\beqa{\begin{Eqnarray}}
\def\eeqa#1{\label{#1}\end{Eqnarray}}
\def\eeqan{\end{Eqnarray}}

\let\bar=\overbar

\def\lsim{\mathrel{\raise.3ex\hbox{$<$\kern-.75em\lower1ex\hbox{$\sim$}}}}
\def\gsim{\mathrel{\raise.3ex\hbox{$>$\kern-.75em\lower1ex\hbox{$\sim$}}}}

\def\del{\partial}
\def\Dslash{\not{\hbox{\kern-4pt $D$}}}
\def\dslash{\not{\hbox{\kern-2pt $\del$}}}
\def\pslash{\not{\hbox{\kern-2pt $p$}}}
\def\ETmiss{\not{\hbox{\kern-4pt $E$}}_T}

\def\Dlr{\mathrel{\raise1.5ex\hbox{$\leftrightarrow$\kern-1em\lower1.5ex\hbox{$D$}}}}

\def\MSB{{\bar{M \kern -2pt S}}}
\def\msb{{\bar{\scriptsize M \kern -1pt S}}}

\def\drb{{\bar{\scriptsize D \kern -1pt R}}}

\newcommand\snowmass{\begin{center}\rule[-0.2in]{\hsize}{0.01in}\\\rule{\hsize}{0.01in}\\
\vskip 0.1in Submitted to the  Proceedings of the US Community Study\\ 
on the Future of Particle Physics (Snowmass 2021)\\ 
\rule{\hsize}{0.01in}\\\rule[+0.2in]{\hsize}{0.01in} \end{center}}

\begin{document}


\Title{RF Electronics}

\medskip 

\Author{Josef Frisch (SLAC), Paul O'Connor (BNL)}

\medskip


\medskip

 \begin{Abstract}
\noindent 
For many decades High Energy Physics (HEP) instrumentation has been concentrated on detectors of ionizing radiation -- where the energy of incident particles or photons is sufficient to create mobile charge in gas, liquid, or solid material, which can be processed by front end electronics (FEE) to provide information about the position, energy, and timing of the incoming radiation. However, recently-proposed  HEP experiments need to sense or control EM radiation in the radiofrequency (RF) range, where ionization detectors are unavailable. These experiments can take advantage of emerging microelectronics developments fostered by the explosive growth of wireless data communications in the commercial sector.

Moore's Law advances in semiconductor technology have brought about the recent development of advanced microelectronic components with groundbreaking levels of analog-digital integration and processing speed. In particular, RF "System-on-Chip" (RFSoC) platforms offer multiple data converter interfaces to the analog world (ADCs and DACs) having bandwidths approaching 10 GHz and abundant digital signal processing resources on the same silicon die. Such devices eliminate the complex PC board interfaces that have long been used to couple discrete ADCs and DACs to FPGA processors, thus radically reducing power consumption, impedance mismatch, and footprint area, while allowing analog preconditioning circuits to be eliminated in favor of digital signal processing. Costed for wide deployment, these devices are helping to accelerate the trend towards “software defined radio” in several high-volume commercial markets. In this whitepaper we highlight some HEP applications where leading-edge RF microelectronics can be a key enabler.
\end{Abstract}

\snowmass

\def\thefootnote{\fnsymbol{footnote}}
\setcounter{footnote}{0}

\section{Introduction}
In recent years the High Energy Physics community's experimental portfolio has begun to require instrumentation to probe the classical properties of the electromagnetic field -- the spatial and temporal behavior of amplitude, phase, frequency, and polarization -- with unprecedented signal-to-noise. These experiments are now able to take advantage of developments in semiconductor microelectronics brought about by the growth of commercial wireless applications, particularly telecommunications, radar, and wireless powering. Modern microelectronic platforms have now reached the stage where most signal processing functions conventionally realized with analog circuits can be subsumed into the digital fabric of Field Programmable Gate Arrays which incorporate RF-rate data converters with large DSP "slices", memories, I/O structures, and processor cores. For many systems this leads to orders-of-magnitude improvement in signal processing throughput per unit of power consumption. In the following sections we will describe some of the HEP activities making use of the new digital RF technologies and projections for their use in future large-scale programs. In many cases, citations to the relevant proposals in other Snowmass whitepapers are provided.

\begin{wrapfigure}{l}{0.5\textwidth}
  \vspace{-20pt}
  \begin{center}
    \includegraphics[width=0.48\textwidth]{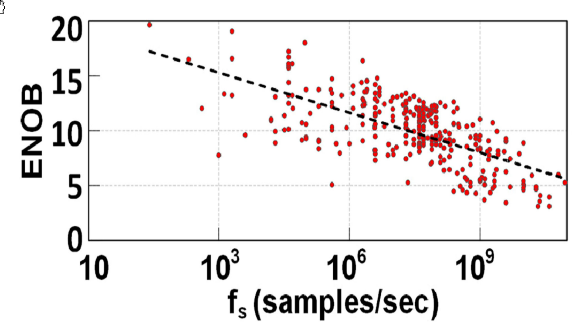}
  \end{center}
  \vspace{-20pt}
  \caption{Analog-to-digital converter resolution (effective bits) vs. sampling rate  c. 2016}
  \vspace{-10pt}
  \label{fig:adcENOB}
\end{wrapfigure}

Prior to 2018, two bottlenecks stood in the way of fully-digital processing of high bandwidth, high dynamic range signals. First, ADC performance (as measured on the speed/resolution plane, see Figure) was barely adequate to acquire multi-GHz signals at Nyquist rate, having only a few bits of resolution (Figure 1). Therefore, heterodyne techniques involving costly, sensitive analog preconditioning circuitry (mixers, local oscillators, quadrature couplers, etc.) were used to translate down to intermediate frequencies for lower-rate digitization. Second, the interface that transfers digitized data between the high-speed ADC and first-stage digital processor became complex and power-consuming, and required expert circuit layout techniques to avoid timing errors. The cost and design effort to realize high-bandwidth, high-resolution digital processors relegated them to specialty markets.

\begin{wrapfigure}{l}{0.5\textwidth}
  \vspace{-20pt}
  \begin{center}
    \includegraphics[width=0.48\textwidth]{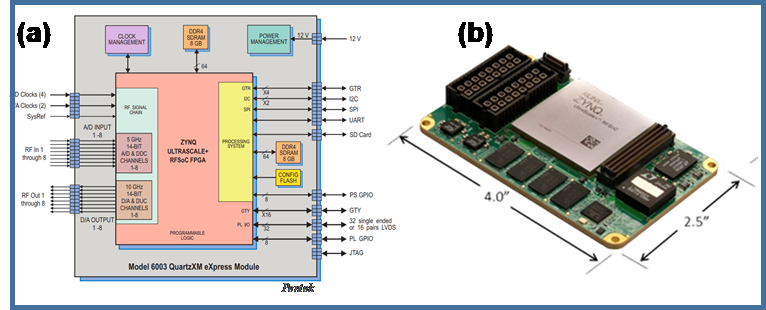}
  \end{center}
  \vspace{-20pt}
  \caption{(a) block diagram and (b) form factor for an RFSoC module offered by Pentek}
  \vspace{-10pt}
  \label{fig:RFSoC}
\end{wrapfigure}

An emerging family of microelectronic platforms dedicated to acquiring and processing analog RF signals in the digital domain has been released by Xilinx starting in 2018. Referred to as Zynq UltraScale+ RFSoC (hereafter RFSoC) they incorporate high-rate analog-to-digital converters (ADCs) along with a rich fabric of fast programmable logic and memory, powerful quad-core ARM processors for application processing, and abundant high-speed connectivity interfaces. The ADCs have a combined sample-rate/dynamic range/power consumption figure of merit exceeding that of the most advanced discrete devices, and by providing a built-in, on-chip interface between digitizers and signal processing logic, design firmware development cycles can be shortened by an order of magnitude. Several vendors now offer the RFSoC chips on compact form factor modules with turnkey base firmware and a variety of custom IP for typical applications. Target markets include new high-bandwidth wireless networking and telecommunication, radar, satellite communication, and aerospace/defense applications. Xilinx has introduced three generations of RFSoC between 2018-2020, each time increasing the maximum ADC sampling rate by 25\%, and further improvements are expected as new sub-10nm CMOS process nodes become available. 

Interestingly, the same technical requirements demanded by commercial product developers overlap significantly with research areas of interest to the community:
\begin{itemize}
    \item multiple, phase-coherent sampling channels;
    \item analog bandwidth \textgreater 5GHz with sampling rate \textgreater 2Gsa/s and resolution \textgreater 10 effective bits;
    \item low noise, distortion, and crosstalk;
    \item moderate power consumption and compact form factor;
    \item configurable, field-programmable functionality;
    \item scalable to large systems.
\end{itemize}

\begin{wrapfigure}{r}{0.5\textwidth}
  \vspace{-20pt}
  \begin{center}
    \includegraphics[width=0.48\textwidth]{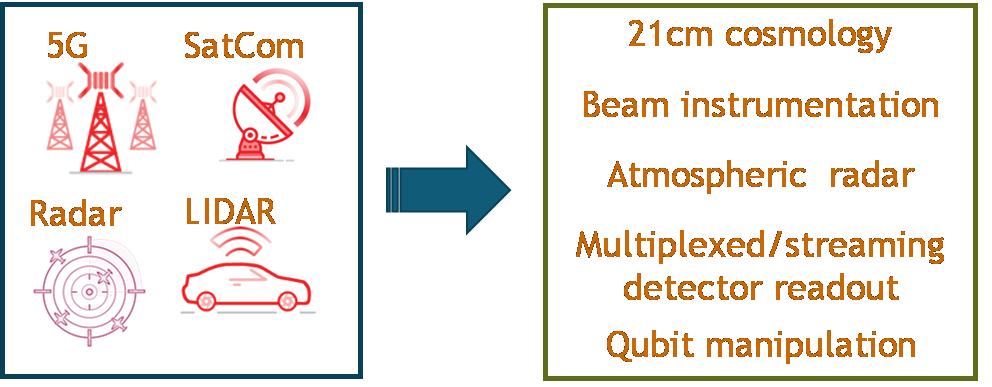}
  \end{center}
  \vspace{-20pt}
  \caption{commercial application requirements map onto experimental needs.}
  \vspace{-10pt}
\end{wrapfigure}


The benefits of integrated RF platforms are beginning to be realized in the experimental physics community. Recent activities include accelerator applications \cite{dusatko2019evaluation}, radio astronomy \cite{liu2021characterizing}, atmospheric sensing \cite{conway2019element}, multiplexed readout of superconducting bolometers \cite{sinclair2020development} and scintillators \cite{mishra2021recovery}, radio neutrino detection \cite{huege2017radio}, and manipulation of qubit structures for Quantum Information Science \cite{borgani2021adapting}.


\section{Microwave multiplexing for CMB}
Cryogenic bolometers are widely used for Cosmic Microwave Backgrond (CMB) intensity and polarization mapping.  Detectors have noise levels below the CMB shot noise so recent developments have focused on increasing the number of detector elements in a telescope system.  The Simons Observatory is in production of a readout with ~70,000 detectors, and future projects will use hundreds of thousands of detectors

\subsection{Cryogenic Bolometer Technology}
Detectors for CMB are commonly operated at ~100mK temperatures in order to have detection noise that is small compared to the CMB shot noise and sky noise. The most commonly used type of detector for CMB instruments is the Transition Edge Sensor (TES). These are fabricated as thin films composed alloys with a superconducting transition temperature slightly above the cryogenic bath temperature.  When voltage biased, a natural electro-thermal feedback maintains the detector at its transition temperature, and the device current is dirctly realted to external power applied to the detector. 

The TES output current requires some architecture of SQUID amplification before it can be further amplified by conventional semiconductor amplifiers. In existing telescope systems this amplification is combined with multiplexing. Three types of systems are in common use:

\begin{itemize}
    \item Time division Multiplexing (TDM): This is the most commonly used scheme for CMB. In this scheme a row / column address mechanism is used to reduce the number of readout wires. This provides O(100) multiplexing ratios
    \item MHz Frequency Division Multiplexing (FDM): This scheme incorporates the TES sensors in MHz frequency resonant circuits, with each resonator tuned to a different frequency.  This provides O(100) multiplexing ratios
    \item Microwave Multiplexing (uMUX0:  This scheme couples the SQUID current to SQUID loaded microwave resonators, typically operating in the 4-8GHz frequency range, with each resonator tuned to a different frequency.  The SQUIDs act as a type of parametric amplifier allowing the low power TES current to modulate a  higher power RF signal.  This provides O(1000) multiplexing ratios. The high multiplexing ratios and excellent noise performance of these systems is attractive, but they require complex RF based warm readout systems.
\end{itemize}

Kinetic Induction Detectors (KIDs) make use of the temperature dependence of the inductance of superconducting circuits operated near their transition temperatures. When fabricated into resonant circuits, this provides a temperature tuned resonant circuit that can be used to frequency multiplex many detectors.  KID detectors have not yet matched the noise performance of TES detectors, however the technology is improving and the potential multiplexing factors of O(10,000) make them attractive for future CMB systems.   The operating frequencies range from a few X 100MHz to several GHz.  The readout systems for KIDs are very similar to microwave multiplexing readouts.

\begin{figure}[H]
\begin{center}
    \includegraphics[width=\textwidth]{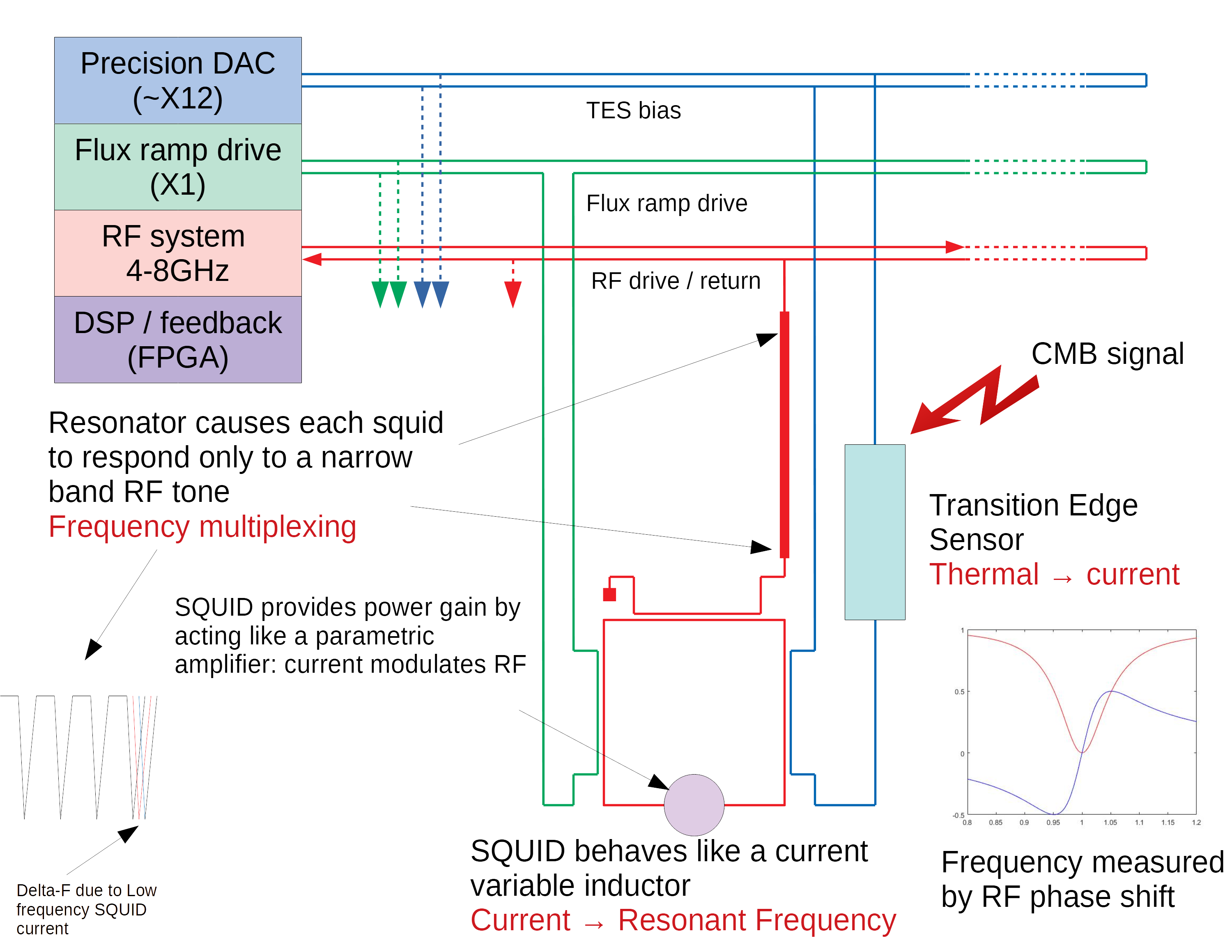}  
    \caption{Warm readout system for microwave multiplexed TES detectors}
\end{center}
\label{fig:cryodet}
\end{figure}

\subsection{Warm readout System}

uMUX systems and KID systems use nearly identical microwave readout systems. These systems much generate RF tones to interrogate the cold resonators in order to measure the resonator frequency, and then tune those tones to track the frequency changes of those resonators. This comprises a digital radio system operating with O(1000) tones for uMUX and potentially O(10,000) tones for KID detectors. The "warm" readout can be considered to comprise all of the "conventional" RF devices from the 4K amplifiers to the digital feedback system.  The warm readout system has a number of technical challenges:
\begin{itemize}
    \item Bandwidth: The readout system typically requires 4GHz of analog bandwidth. This exceeds the anti-aliased bandwidth of high resolution ADCs / DACs as of 2022.  Note that while in principal an ADC has a Nyquist bandwidth of 1/2 of the clock speed, in practice antialias filters have finite slopes, and a ration of 3:1 to 5:1 of digitizing rate / analog bandwidth is typically required. This necessitates the use of multiple ADCs and DACs to covere the 4GHz bandwidth. These can be combined using frequency multiplexing filters, or in an I/Q configuration.
    \item Signal Processing:  The total digital bandwidth is large, as an example the SLAC SMuRF uMUX system uses 8X 2.5Gs/s digitizers to obtain 4GHz of bandwith, resulting in a ~400 Gbit raw data rate.  Processing throughput is near the limit of modern FGPAs using efficient code and algorithms.
    \item Dynamic range: Any RF system will have some nonlinear response.  This results in the production of inter-modulation signals. 2nd order signals have frequencies of f1+f2 and f2-f2.   In some designs (such as SLAC SMuRF these tones are outsdie of the signal band, however that required the use of a larger total digitizing throughput.   The 3rd order terms of the form F1+F2-F3 will be in band and the number of those tones scales as N\textsuperscript{3}  With O[10\textsuperscript{3}]tones this results in O[10\textsuperscript{9}] intermodulation tones.  The large number of intermodulation tones behave like noise floor.
    \item Tone tracking: For both uMUX and KID detectors, the frequency modulation from signals (and flux ramp for uMUX) is not small compared to the resonator line width.  Having the drive tones track the resonator frequency shifts provides improved linearity with respect to signal inputs and reduces the signal levels in the RF chain, improving the RF dynamic range. Tone tracking requires additional firmware complexity and additional FPGA processing power
\end{itemize}

The present generation of uMUX readout systems are capable of reading approximately 2000 detectors in a 4GHz bandwidth, and require approximately 1U of rack space and 250W of power. The next generation of readout based on RFSoC parts is expected to reduce the size and power consumption by a factor of 2X-4X. 

Note that in addition to the RF systems uMUX readouts require a significant amount of high precision, low frequency systems for TES bias and other functions. The underlying industrial technology is not developing as quickly as is the RF / signal processing and is expected to represent a larger fraction of the total system size and complexity in the future. KID detectors have a substantial advantage here in not requiring the large low frequency bias system. 

\section{Instrumentation for 21cm cosmology}

\begin{wrapfigure}{L}{0.5\textwidth}

  \begin{center}
  \vspace{-20pt}
       \includegraphics[width=2.5in]{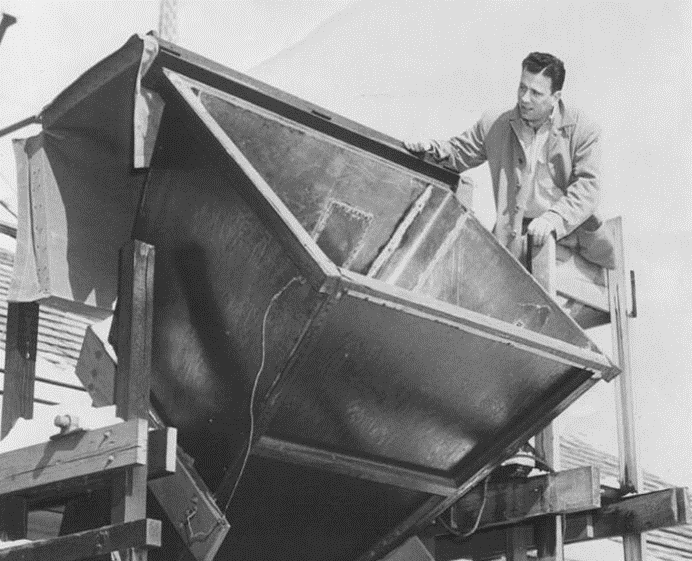}
   \end{center}
   \vspace{-20pt}
\caption{\footnotesize Harold Ewen inspects the horn of the instrument he used to detect the Galactic 21\.cm radiation. Credit: NRAO/AUI/NSF}
\vspace{-10pt}
\label{fig:ewen_figure}
\end{wrapfigure}

The 21\,cm hyperfine transition of neutral atomic hydrogen (HI) occurs when the electron and proton spins change from aligned to anti-aligned. Observation of the radiofrequency emission from the cosmos began in 1951 \cite{ewen1951observation} using instruments made possible by the development of radar during World War II. It was soon used to map the structure and rotation curve of our galaxy and later, neighboring galaxies -- which has provided the strongest evidence for the existence of dark matter. 

There are many  instrumentation challenges to extending 21cm observations to cosmological distances, but the potential to improve our understanding of fundamental processes underlying cosmic acceleration (dark energy and inflation) is great. Since the 21\,cm transition is sharp and isolated, we can use it to determine precise source redshifts. This opens the possibility to probe regions of the Universe that are inaccessible to optical surveys. However,  the amplitude of the 21\,cm signal is very weak and is masked by strong foregrounds from galactic synchrotron and point source emission, as well as anthropogenic RFI, in the same frequency bands. 

Two lines of R\&D have emerged over the past decade and progress is being made in overcoming the main issues. The first, targeting large-scale structure in the modern universe at $0.1 \leq z \leq 6$, employs \emph{intensity mapping} to detect the aggregate emission from many hundreds of unresolved galaxies using telescopes whose angular and frequency resolution are coarse, but sufficient to map structures of relevance for cosmology. There has been an initial detection in cross-correlation with optical galaxy surveys at low redshift using the 100-m dish of the Green Bank Telescope \cite{masui2013measurement} in 2013. Follow-on experiments such as CHIME \cite{newburgh2015measuring} using large interferometers are underway with greater sensitivity. The second approach seeks to characterize the global (monopole) signal from neutral hydrogen during the cosmic Dark Ages ($10 \leq z \leq 500$), an epoch in cosmic history that is otherwise unobservable as stars and galaxies had not yet formed. During this time cosmological processes are unaffected by astrophysical interactions and provides a pristine laboratory which has shown tantalizing signs of BSM physics \cite{Bower2019}.

\subsection{HI Intensity Mapping}
By virtue of its inherently large field of view and high survey speed, intensity mapping interferometers will likely surpass optical surveys in terms of volume of the cosmos surveyed. Achieving this potential will require a new generation of powerful, scalable on-antenna receivers and digitizers based on modern RF microelectronic platforms; their individual-element data streams will be phase-synchronized and transported to a massive digital correlator with processing throughput equalling or exceeding that of present-day supercomputers (see Fig. \ref{fig:data_flow_block_diagram}). 

High channel-count, fast RF digitization as planned for large 21\,cm interferometers is also used extensively in accelerator control and beam diagnostics. Large accelerators such as LCLS-II can include over a thousand channels of RF front ends and high-performance digitizers connected to a distributed data network. Although optimization of dynamic range, bandwidth, and noise characteristics differ from those needed for the 21 cm experiment, many commonalities between the designs remain. For performing early digitization at every antenna or antenna group, the RF system-on-chip (RFSoC) family which integrates GHz-rate data converters with powerful digital logic, DSP, and memory \cite{liu2021characterizing} is ideally suited. Current generation RFSoC parts have up to eight, 6 Gigasample/sec converters with 12-bit nominal resolution (11.4 ENOB). This allows a tradeoff: high bandwidth direct digitization provides the ability to oversample and design high performance digital band selection filters and high order frequency channelizers, relaxing constraints on the analog prefiltering and conditioning. The onboard programmable logic will also permit field-programmable algorithms to perform calibration and RFI flagging and mitigation. These are well understood problems in radio astronomy, though the effects of residual RFI at the small level of the 21\.cm signal is only starting to be addressed \cite{harper2018potential}.

Current generation 21\,cm instruments produce $\ge$100TB of data per day without any compression, natively generating an amount of data $\propto N^{2}$ where $N$ is the number of elements (currently $N \sim 10^{3}$). Compression by a factor of $\sim N$ is achievable by exploiting redundancy within the interferometer, but requires the use of real-time, in-situ calibration and places strong constraints on the uniformity of the optics beteen elements. Estimates of data rates, correlator processing speeds, and compute power for a proposed large  ultrawideband IM experiment (PUMA \cite{bandura2019packed}) are given in Ref. \cite{castorina2020packed}. The scale of the problem can be seen by comparison with present large data acquisition systems: the rates estimated for PUMA exceed those at the LHC, LCLS-II, and SKA1 by factors of over 100 up to several 1000. One approach being studied to constrain the enormous power dissipation is a dedicated ASIC having sufficient numbers of complex multiply-accumulate units (CMACs) to allow all correlations of up to 64 signals to be computed in parallel at an energy cost of as low as 1.8 pJ per CMAC operation \cite{d2016low}.

\begin{figure}[htb!]
  \centering
  \includegraphics[width=5in]{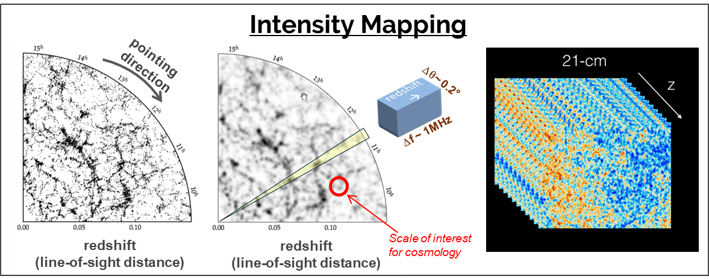}
\caption{\footnotesize HI intensity mapping (IM) makes a tomographic reconstruction of large-scale structure by capturing the aggregate emission of many unresolved galaxies. Left: Distribution of galaxies in 2D (radial coordinate is distance) from an optical galaxy spectrographic survey. Middle: Illustrating the resolution in angle and frequency relevant to cosmology. Right: Map-making results in a density map at each redshift.}
\label{fig:IM_figure}
\end{figure}

\begin{figure}[htb!]
  \centering
  \includegraphics[width=5in]{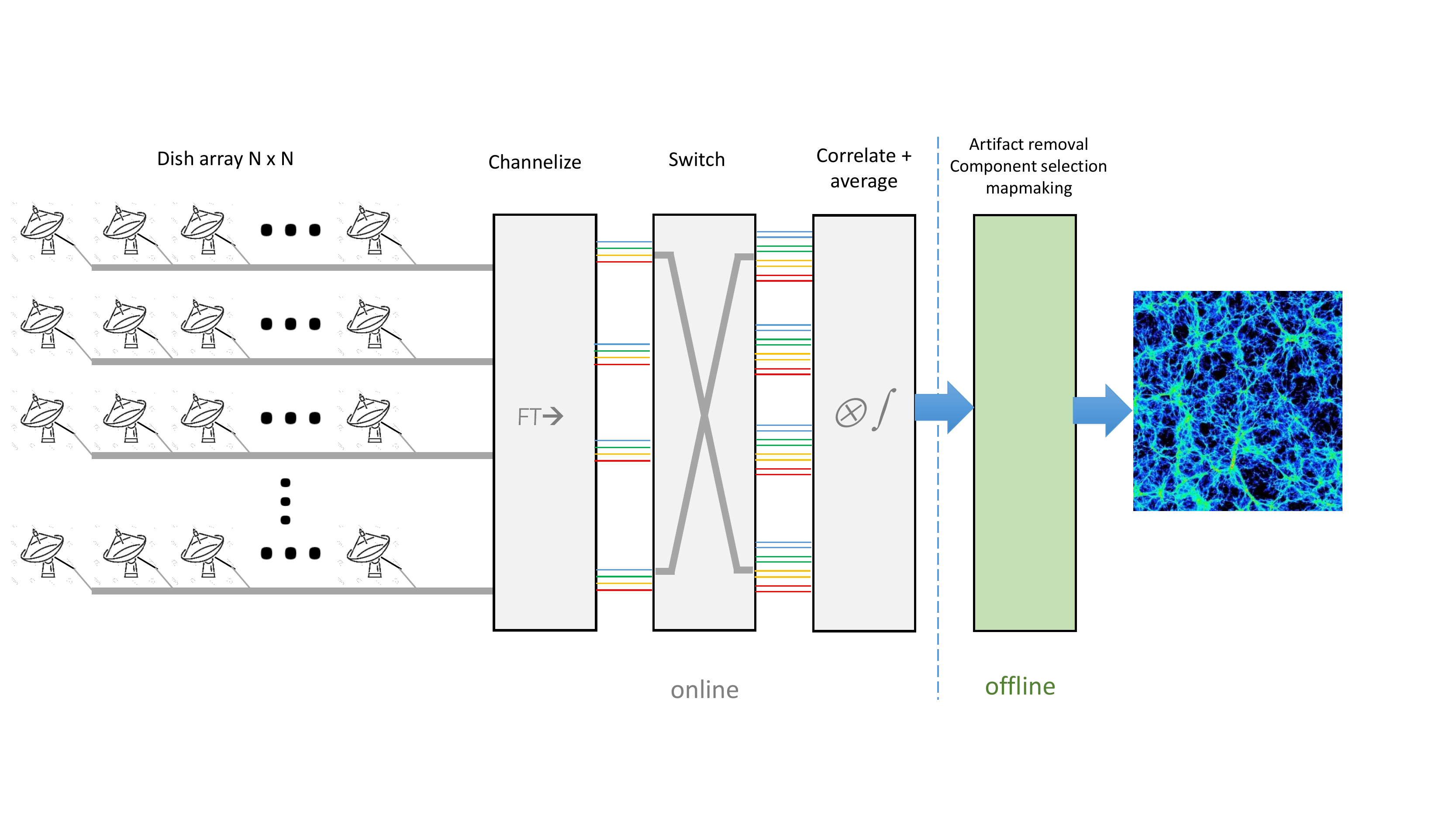}
\caption{\footnotesize Illustration of anticipated data flow in a large interferometric array.  Conversion of waveform data to frequency space, e.g. channelization, is accomplished close to each receiver; coincident data for each frequency bin are collected from all stations through a cross-bar switch (also called a ``corner-turn'' operation); correlations are constructed for each frequency bin, which can then be time-averaged and stored, followed by physics analysis.}
\label{fig:data_flow_block_diagram}
\end{figure}

\subsection{Measuring the monopole signal at high redshift}
The redshift range $10 \leq z \leq 500$ is a largely unexplored window on the infant Universe. During the so-called Dark Ages the Universe was dominated by neutral atomic hydrogen and the population of the upper and lower states of the 21\,cm system dictates whether the signal is seen in emission or absorption against the CMB backlight. Observing the global (sky-averaged) 21\,cm signal provides a measure of the interplay between the HI spin temperature and cooling due to cosmic expansion. The absence of astrophysical effects in the early Universe makes robust theoretical predictions possible, and any departure from standard cosmological models is an indicator of new physics. 

At frequencies below $\sim$ 10 MHz the absorption, emission, and distortion of the Earth's ionosphere prevent ground-based measurements except at the most favorable sites during ideal conditions. In Earth orbit, above the ionosphere, extensive terrestrial RFI (from man-made or meteorological sources) similarly prohibits meaningful observations in the HF (3 - 30 MHz) and UHF (30 - 300MHz) bands. For decades it has been recognized that   the lunar farside, where neither terrestrial RFI nor ionospheric effects are present, presents an ideal site for low frequency radio astronomy \cite{basler1976evaluation}, \cite{johnson1985siting}. The potential of the 21\,cm probe has been recognized by the NASEM Decadal Survey of Astronomy and Astrophysics, whose Panel on Cosmology identified the Dark Ages as its sole discovery area. A number of Snowmass Cosmic Frontier whitepapers (\cite{Liu2019Cosmology}, \cite{liu2022snowmass2021}, \cite{chakrabarti2022snowmass2021}, \cite{boddy2022astrophysical}) develop the science case for a 21\,cm Dark Ages experiment on the Moon. An alternative, equally viable approach is to observe from a satellite or satellite formation in a lunar orbit selected to provide periodic blocking of terrestrial RFI, wider and higher-cadence sky coverage, and interferometric imaging (\cite{shi2022lunar}, \cite{shi2022imaging}). 

RF electronics with the best possible balance of processing power for a highly constrained mass and power budget will be the key enabling technology for a lunar Dark Ages experiment, along with optimized antennas. In the short term only modest missions with shared instrument payloads will be possible for cost reasons. The frequency range is  low compared to modern-Universe IM experiments and digitizers in the $\sim$ 100 Msamples/sec are appropriate. The analog signal chain requires careful attention to the interface between the antenna and preamplifier to achieve a noise level low enough to be sensitive to the Dark Ages signal in the presence of foregrounds nearly $10^5$ times larger. The only possibility for a small pathfinder mission to detect the signal is by integrating over many lunar cycles, and by appropriate foreground modeling along with accurate antenna beam characterization to allow deconvolution. The receiver analog signal chain must have high  dynamic range to handle the sky signal as well as any unmitigated EMI from the rest of the spacecraft.

The harsh lunar environment presents significant challenges for the instrument package: radiation dose is about 2.5$\times$ what is encountered in low Earth orbit, and the surface temperature ranges from $\sim$ 100K to $\sim$ 400K between over the 28-day lunar cycle. Unless radioisotope thermal generators are used, the instrument electronics will need to operate on battery power for the duration of the lunar night (14 Earth days). With mass budgets in the 100kg range, battery capacity will likely limit the nighttime power budget to around 10W. During the lunar day the battery will be recharged by solar photovoltaics. Besides the difficulty of lunar surface survival, a telescope on the farside is necessarily permanently out of direct contact with Earth ground stations. Therefore, radio communication with a relay satellite will be required and with low-power transceivers it will be necessary to use powerful encoding and compression to transport the acquired data.

As in the IM case, calibration methods will need to be developed to allow separation of galactic foregrounds from the cosmological 21\,cm signal. There are no natural calibrators available (with the possible exception of the outer radioplanets), and one proposal is to insert a broadband transmitter into a lunar orbit where it can be viewed by the receiver on the surface. By using coded modulation on the calibrator, it can be possible to detect the signal at amplitudes low enough to not interfere with normal observing. Various schemes for developing the RF circuitry for an orbiting calibrator are being studied.

\section{Radio Detection of UHE neutrinos, cosmic rays, and photons}
\begin{wrapfigure}{l}{0.5\textwidth}
  \vspace{-20pt}
  \begin{center}
    \includegraphics[width=0.48\textwidth]{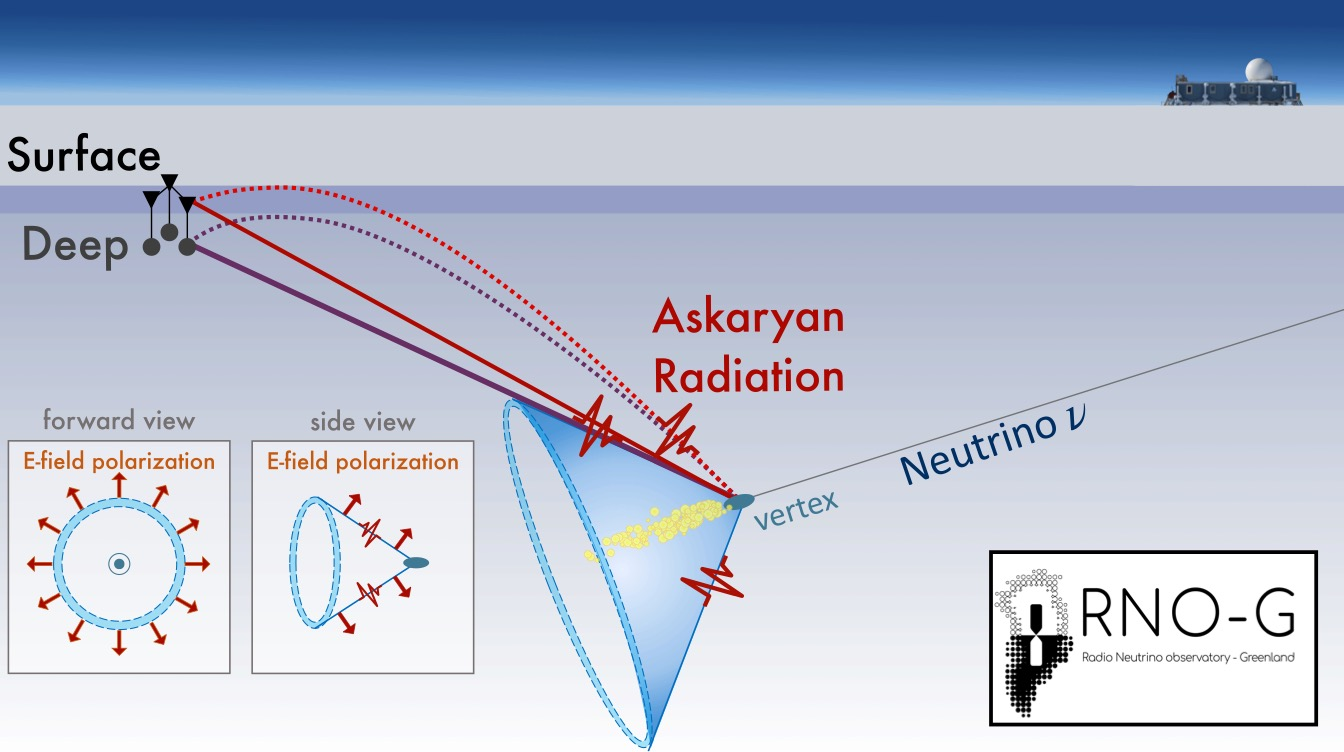}
  \end{center}
  \vspace{-20pt}
  \caption{Radio neutrino detection (credit: events.icecube.wisc.edu)}
   \vspace{-30pt}
  \label{fig:ice_cube}
\end{wrapfigure}

The highest energy neutrinos probe extreme astrophysical environments inaccessible via other messengers \cite{aguilar2021design}. When a high-energy neutrino interacts, a fraction of its energy is transferred through deep-inelastic scattering with nucleons, which causes a hadronic shower to develop. Radio detection of neutrinos is possible through the Askaryan effect (Fig.  \ref{fig:ice_cube}), where particle showers in dense media (in this case polar ice) cause nanosecond-scale radio pulses in the frequency range between 30 MHz and 1 GHz.  Consequently, fast, broad-band and low-noise receivers and systems are needed to efficiently detect the rare signals. Dozens to hundreds of such receivers in an interferometric phased array distributed over a few square km are being proposed, requiring cost- and power-efficient modules having exquisite (sub-nanosecond) timing precision to accurately reconstruct event topologies. Experiments such as ARA \cite{seikh2022towards}, ANITA \cite{gorham2019constraints}, and ARIANNA \cite{gorham2019constraints} have begun studying the diffuse flux of astrophysical neutrinos and candidate sources of extra-galactic neutrinos. Proposed new in-ice facilities such as RNO-G \cite{aguilar2021design} and IceCube-Gen2 \cite{blaufuss2016icecube} will monitor volumes of up to several 100 $km^{3}$ of polar ice with the goal of detecting the first EeV neutrinos. These experiments call for large arrays of radio receivers with sensitivity over a wide range of frequency to enable reconstruction of event energy and topology. Leveraging the R\&D directed at commercial markets will be critical to cost-efficient construction of large facilities for UHE particle astrophysics in the coming decade.

In addition to neutrinos, the performance of large-scale air-shower arrays can be enhanced by augmentation with a radio sub-detector, as the radio emission in air showers is sensitive to the electromagnetic component. A hybrid detector of this type has been shown \cite{Renschler_2019} to improve sensitivity to highly-inclined extensive air showers such as those resulting from PeV gamma rays originating from the Galactic Center.

\vspace{20pt}
\section{Other applications}
\subsection{Axion dark matter searches}
Laboratory experiments to search for "wavelike" dark matter seek to detect the conversion of axions to photons in a high-Q resonant cavity in the presence of a magnetic field. If this conversion can occur, then its rate will be enhanced when the cavity's resonant frequency matches that of the produced photon, which corresponds to the mass-energy of the axion. Thus far searches have not resulted in detection, but future experiments will extend the frequency range and sensitivity to cover a wider search range. Cavity resonant frequencies cover a wide range in the microwave spectrum and require extensive processing before and after digitization to extract the weak signals from axion conversion. Modern RF microelectronic technology will play a key role in carrying out the next generation of experiments.

\subsection{Control and manipulation of superconducting qubit systems}

Superconducting qubit systems offer a promising way to work with information encoded in uniquely quantum properties (entanglement and coherence). Such systems behave like macroscopic  "artificial atoms" whose energy band structure corresponds to transition frequencies in the microwave range. Typically, warm electronic modules generate and detect the necessary analog  signals that operate on the qubit system to create, sequence, and probe system states. The generation and transmission of control signals is challenging because small errors in control signals will affect the results of operations, and the errors associated with each gate operation accumulate as the system runs. As an example of how modern RF technologies developed for wireless telecom has already been put to use for quantum information processing, see Reference \cite{borgani2021adapting}.








\bibliographystyle{JHEP}
\bibliography{main1}






\end{document}